# Monolithic all-Silicon Flat lens for broadband LWIR Imaging

Orrin Kigner[1], Monjurul Meem[2], Brian Baker[3], Sourangsu Banerji[2], Philip W. C. Hon[1], Berardi Sensale-Rodriguez[2] and Rajesh Menon[2,4]

[1]Northrop Grumman Corporation, NG Next, Redondo Beach CA 90278, USA.
[2]Department of Electrical & Computer Engineering, University of Utah, Salt Lake City UT 84112, USA.
[3]Utah Nanofab, University of Utah, Salt Lake City UT 84112, USA.
[4]Oblate Optics, Inc. San Diego CA 92130, USA.
*rmenon@eng.utah.edu

**Abstract:** We designed, fabricated and characterized a flat multi-level diffractive lens (MDL) comprised of only silicon with diameter = 15.2mm, focal length of 19mm, and operating over the longwave infrared (LWIR) spectrum of 8μm to 14μm. We experimentally demonstrated field of view of $46^0$, depth of focus >7mm and wavelength-averaged Strehl ratio of 0.46. All these metrics were comparable to those of a conventional refractive lens. The active device thickness is only 8μm and its weight (including the silicon substrate) is less than 0.2g.

Planar lenses can reduce the weight and thickness of imaging systems. Multi-level diffractive lenses (MDLs) are planar lenses where the complex transmission function is computed using inverse design, and subsequently, implemented as a nano- or micro-structured surface [1]. MDLs have been shown to be highly effective at achieving large numerical aperture (NA), [1] bandwidth [2] and depth of focus [3]. Compared to alternatives like metalenses, MDLs potentially offers better performance as well as somewhat easier manufacturing due to their larger minimum feature sizes [4].

Flat lenses in the long wave infra-red (LWIR) wavelength range of 8μm to 14μm have not been studied significantly. Metalens demonstrations so far have been limited to only single wavelengths (λ=10.6μm). These include a 12mm diameter, NA=0.6 metalens [5], micro-metalenses with diameter of 100μm [6], and polarization sensitive metalenses with diameter of 10.7mm [7]. Of these, only the first two metalenses were monolithic in silicon. On the other hand, MDLs have demonstrated achromatic imaging over the entire LWIR band [8]. In this previous work, the MDL was fabricated in a 10μm-thick photoresist layer on top of a silicon substrate. Due to the non-zero absorption of the photoresist in the LWIR band, the overall performance of the MDL was constrained. Here, we utilize aligned lithography and etch processes to demonstrate an all-silicon monolithic MDL that is achromatic over the LWIR band. In addition, we perform a careful comparison of the Si MDL to that of a conventional refractive lens.

The design procedure for MDLs have been extensively described before [1]. Briefly, the inverse problem consists of finding the MDL surface topography that results in maximum focusing efficiency averaged over the spectral band of interest. We utilize a gradient assisted direct-binary search to solve the inverse problem. Scalar diffraction theory is utilized to model beam propagation from the MDL to the focal plane. Since this formulation is piecewise continuous, gradient computations are well behaved. For demonstration purposes, we chose the following parameters for the MDL: diameter = 15.2mm, focal length=19mm, and λ=8μm to 14μm. The diffraction limit for on-axis focusing enforces minimum feature size of 8μm. Finally, we assumed the refractive index of silicon at the longest wavelength (14μm), which provides the height required to achieve > 2π phase shift across the entire band as ~8μm. In order to minimize the number of lithography and etch steps, we chose to discretize the heights into 8 levels. As has been discussed previously, increasing this number will lead to better efficiency [8,9]. Putting all this together, the MDL surface is discretized into concentric rings of width 8μm, and heights varying between 0 and 8μm in steps of 8 as illustrated by the final design in Fig. 1a.

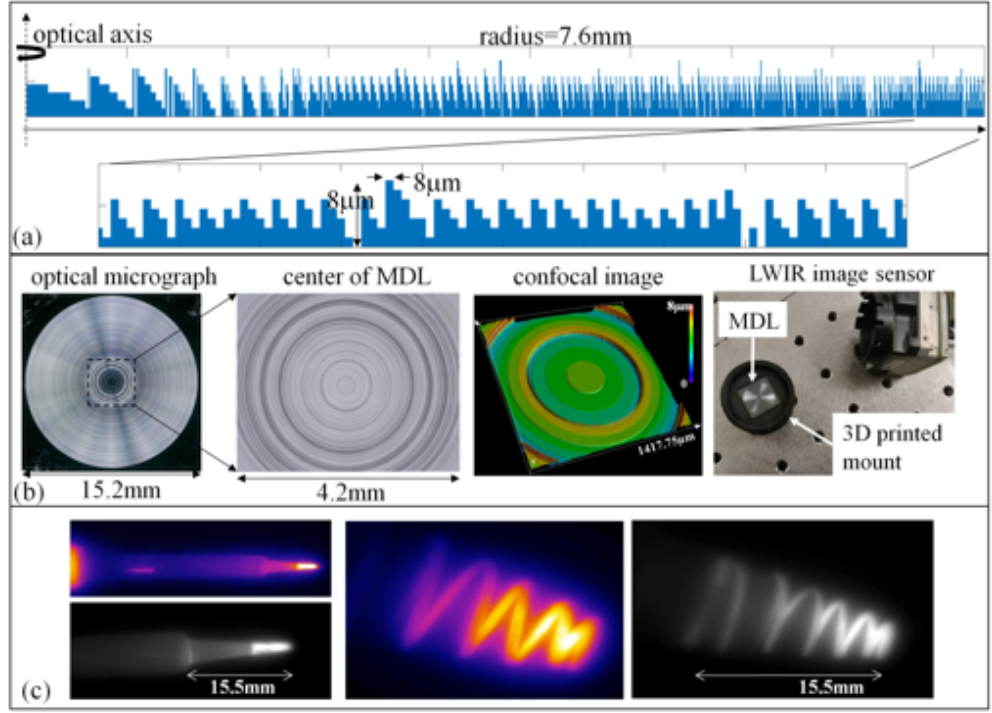

Fig. 1.(a) Radial cross-section of designed MDL. Bottom inset shows the 100 outer rings to emphasize the multi-level features. The optical axis is on the left. (b) Micrographs of the fabricated device. Rightmost panel: Photograph of MDL and LWIR image sensor. (c) Frames from supplementary videos of a soldering iron and coil resistor using the Tau2 (FLIR) sensor and the MDL.

The device was fabricated using optical lithography, followed by a Si etch step (plasma etching with $CF_4$ and $O_2$). The lithography and etch steps were performed 3 times. The pattern was separated into 3 mask layers with overlapping rings. By choosing the etch depths in each mask layer appropriately, it is possible to achieve 8 height levels using 3 lithography steps (see supplementary information) [10]. Micrographs of the fabricated device are shown in Fig. 1b. The weight of the MDL was 0.2g, and this is dominated by the silcon substrate (which has no optical function other than as a mechanical substrate for the patterned 8μm-thick layer). The MDL was mounted in a custom 3D printed holder with C-mount thread to be easily fitted onto an image sensor as shown in the photograph on the right of Fig. 1b. Finally, in Fig. 1c, we show frames from videos acquired using a FLIR Tau2 sensor. The hot objects were placed about 300mm away from the MDL. The simulated PSFs of the MDL as function of wavelength are shown in Fig. 2(a). The PSFs were measured by illuminating the MDL with a quasi-collimated beam from a tunable quantum cascade laser (QCL) (Daylight Solutions MIRcat). A LWIR sensitive focal-plane array (pixel size=17μm, Tau2 FLIR systems) was placed at a distance of 19mm (equal to the focal length) away from the MDL. The recorded PSFs for different incident wavelengths are shown in Fig. 2a. The incident beam was collimated coming out of the QCL and expanded though a 3.3x beam expansion system to fill either the MDL or a conventional ZnSe refractive lens. We confirmed collimation indirectly by characterizing the focused spot size of the conventional ZnSe refractive lens using an Ophir v2 NanoScan beam profiler. Namely, the expansion lenses were adjusted to provide the minimum spot size. The focused spot size was noted to be slightly larger than the theoretical diffraction limited value (measured 16.5μm vs. theoretical 8μm @ λ=8μm). The expansion of the incident Gaussian beam such that it underfills the 1" expansion lens optics is likely the reason for the larger spot size.

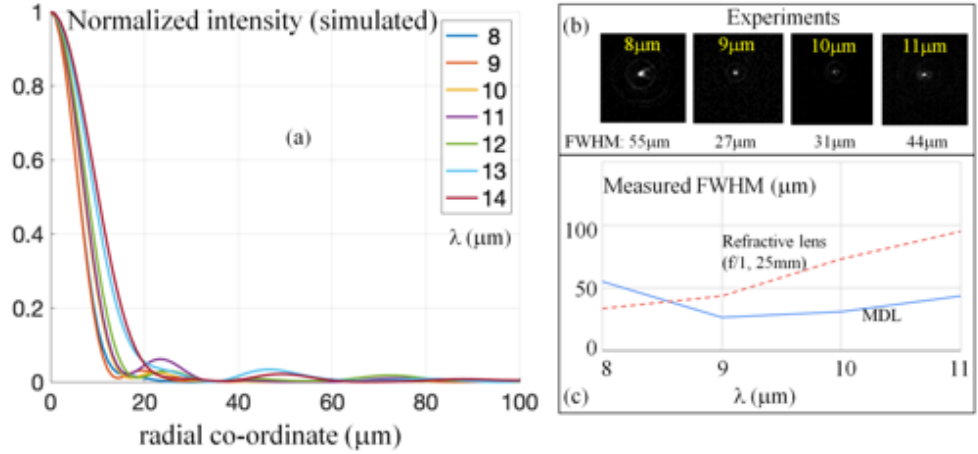

Fig. 2.(a) Simulated PSFs of the MDL in Fig. 1. Good achromatic focusing is observed. (b) Recorded PSFs, and (c) Measured full-width at half-maximum vs λ. The available QCL limited the longest wavelength to 11μm. The dashed line shows the same measurements for a refractive lens.

The same experiments were also performed with a conventional refractive LWIR lens with focal length = 25mm and diameter = 25mm (Thor Labs AL72525-G - Ø1" ZnSe Aspheric Lens). The weight of the refractive lens was 13.1g, more than 5 X higher than that of the MDL. The recorded PSFs are included in the supplementary information. The full-width at half-maximum of the PSFs of the MDL and the conventional refractive lens are plotted in Fig. 2c. The MDL shows smaller FWHM and smaller variation across the spectral band, confirming excellent achromaticity. There has been recent theoretical work that seemed to suggest that the size of flat lenses would limit the operating bandwidth due to the time delay between on-axis and off-axis rays [11]. For our MDL, such a time delay would be 4.9ns. To explore the impact of such a delay, we varied the pulse width of our source from the shortest value accessible to us (20ns) to the longest value (200ns), and captured the two PSFs. The recorded PSFs (see supplementary information) showed no difference, which confirms our hypothesis that such a time delay between the on-axis and off-axis rays are not important, since the integration time of the FPA is many orders of magnitude larger (in our case, this is ~15ms due to frame rate of 30Hz). This test confirms that achieving achromaticity by matching the phase delay between different parts of a planar lens is far too constraining, and is not necessary for achromatic performance.

We also measured the efficiency of the MDL and that of the refractive lens by placing a power meter (1cm x 1cm Gentech Pyroelectric UM9B-BL-D0) before the lens and at the focal plane. Details of the measurement are described in the supplementary information. Since the MDL did not have an anti-reflection coating, its average transmission efficiency (λ=8μm to 11μm) was 23%, compared to 88% for the refractive lens (which does have an anti-reflection coating). The ratio of power measured (area of power meter = 1cm X 1cm) in the focal plane to that measured immediately after the lens was 75% and 99.5% for the MDL and the conventional lens, respectively. Most importantly, the variation in efficiency vs wavelength was minimal (variation < 8% for MDL and <5% for the refractive asphere). The Strehl ratio was extracted from the measured PSFs (see supplementary information) and the variation of Strehl ratio across the LWIR band was minimal, with an average value of 0.46.

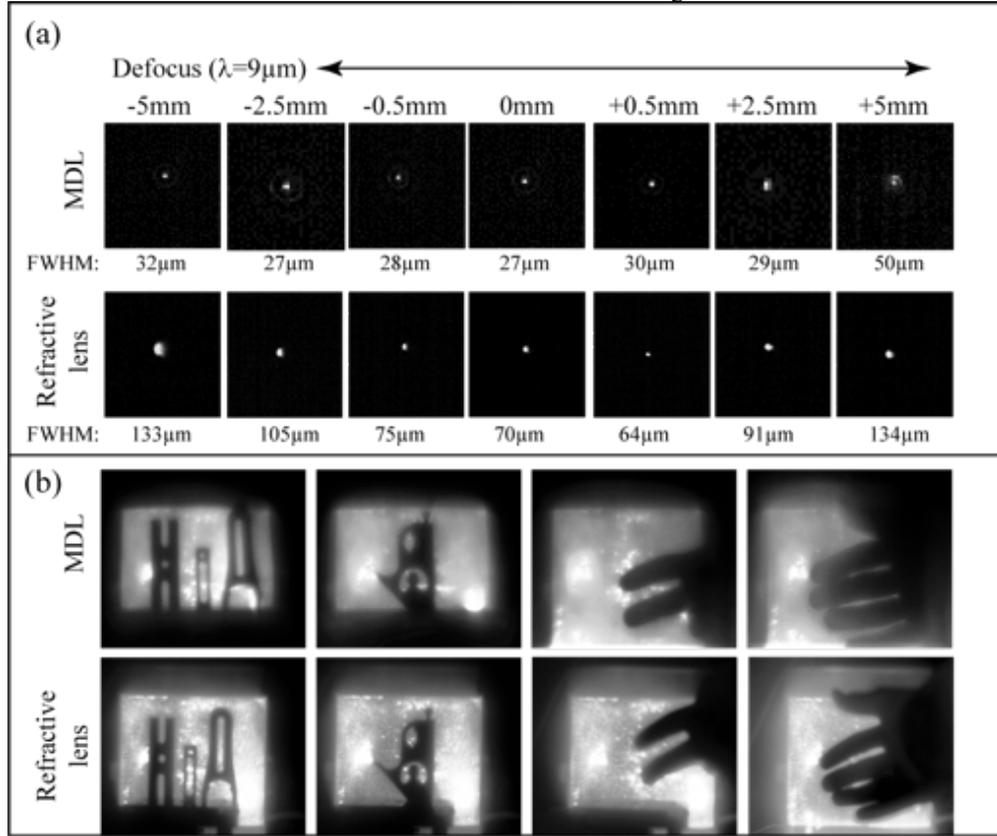

Fig. 3.(a) Measured through-focus PSFs at λ=9μm for the MDL (top row) and the refractive lens (bottom row). The measured FWHM are noted below each image. The depth of focus of the MDL is > 7mm, far larger than that of the refractive lens. (b) Images of various objects backlit by a hot-plate captured by the MDL (top row) and the refractive lens (bottom row). The lower focusing efficiency of the MDL (due to the 8 height levels chosen here) causes additional stray light, which needs to be improved in the future.

We also measured the through-focus PSFs by scanning the distance between the lens and the FPA, and the recorded PSFs are summarized in Fig. 3a. The top and bottom rows show recorded PSFs for the MDL and the conventional lens, respectively. The FWHMs are noted below each PSF. Over a defocus of -5mm to +2.5mm, the FWHM of the MDL changes by only ≤ 3μm, while that of the refractive lens changes by ≤ 63μm. Since the at-focus FWHM for the MDL is 27μm, we estimate its depth of focus to be larger than 7mm. Finally, we also imaged various objects (backlit by a hot plate at $320^0$C). The object distance was between 255mm (images of hand) to 306mm (others), while the image distance was ~19mm. These images are summarized in Fig. 3b. A simple window averaging over 5 pixels was performed in matlab (supplementary information). The corresponding images taken with the refractive lens are shown in the bottom row, and exhibit higher contrast. We attribute the loss in contrast to the lower focusing efficiency of the MDL, which contributes to stray light. This can be mitigated in the future by allowing for more than 8 height levels in the design as discussed before at the expense of increased fabrication complexity [8]. The field of view of the MDL was imaged by imaging a hot plate at various distances from the optical axis (see supplementary information) and we estimated the total FOV to be $46^0$, which was comparable to that of the refractive lens.

In conclusion, in contrast to prior demonstrations of LWIR flat lenses, here we demonstrate an all-silicon multi-level diffractive lens (MDL) that is achromatic over the LWIR band. A second important contribution of this work is the direct comparison to a refractive lens. We conclude that, except for stray light, the MDL performs equal to or better than the comparable refractive lens. Stray light is directly affected by focusing efficiency, which can be improved by increasing the number of design heights at the expense of fabrication complexity.

**Funding.** Office of Naval Research grant N66001-10-1-4065, N000141512316, and DURIP grant: N000141912458. US National Science Foundation ECCS # 1936729.

**Acknowledgments.** A. Majumdar for assistance with thermal imaging and setup.

**Data availability.** Data underlying the results presented in this paper are not publicly available at this time but may be obtained from the authors upon reasonable request.

**Supplemental document.** See Supplement 1 for supporting content.

# Supplementary Information for

# Monolithic all-Silicon Flat lens for broadband LWIR Imaging

ORRIN KIGNER[1], MONJURUL MEEM[2], BRIAN BAKER[3], SOURANGSU BANERJI[2], PHILIP W. C. HON[1], BERARDI SENSALE-RODRIGUEZ[2] AND RAJESH MENON[2,4]

[1]*Northrop Grumman Corporation, NG Next, Redondo Beach CA 90278, USA.*
[2]*Department of Electrical & Computer Engineering, University of Utah, Salt Lake City UT 84112, USA.*
[3]*Utah Nanofab, University of Utah, Salt Lake City UT 84112, USA.*
[4]*Oblate Optics, Inc. San Diego CA 92130, USA.*
*\*rmenon@eng.utah.edu*

1. **Fabrication steps**

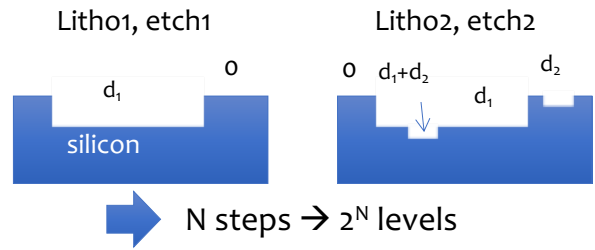

Fig. S1: Multiple lithography and etch process.

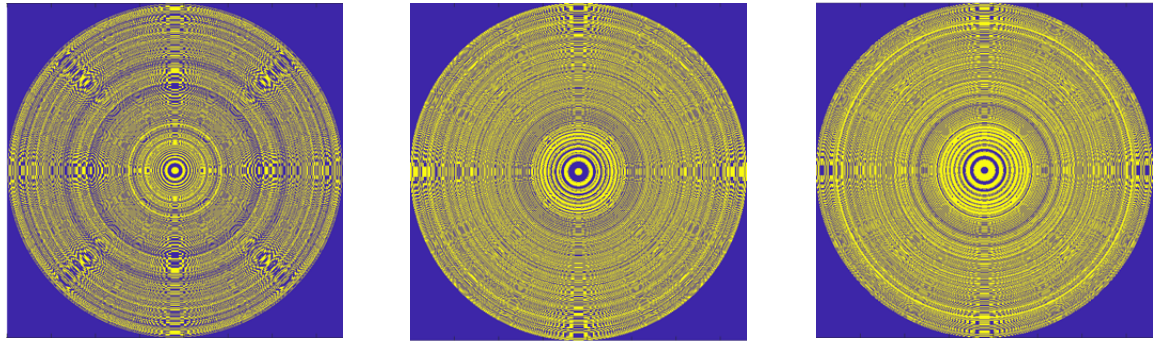

Mask 0
Etch = 8um/7

Mask 1
Etch = 2*8um/7

Mask 2
Etch = 4*8um/7

Fig. S2: The 3 masks used for the MDL in this paper.

A four-inch diameter, n-type polished silicon wafer was cleaned in a two-minute, 200 W oxygen plasma. The wafer was spin-coated with an HMDS adhesion layer, followed by a ten-micron thick layer of AZ 9260 photoresist. After a one-minute, 110-degree Celsius prebake, a chrome-on-glass photomask was aligned and placed in hard contact with the wafer during UV exposure to photolithographically transfer the desired pattern for each etch step into the photoresist. The overlay tolerance of the contact aligner is +/- 1.5 microns. The photoresist was then developed in AZ 1:1 developer solution, after which the silicon area not covered by the photoresist was etched in an Oxford Plasmalab 80 Plus RIE etch system. The etch process employed an etch power of 200 W, a pressure of 10 mT, and a gas mixture of $CF_4$ and $O_2$ gases with flow rates of 26 sccm and 2.6 sccm, respectively. The etch rate was approximately 160 nanometers per minute, and the silicon etching time was proportional to the desired etch depth for each mask step: 1.143, 2.286, and 4.571 microns. The etch uniformity across the wafer was +/- 10 percent. Following the silicon etching, the photoresist was stripped and the process was repeated for each subsequent etch depth. Once wafer processing was complete, the wafer was diced into individual lenses using a Disco DAD641 dicing saw, with outer chip dimensions of 18.1 mm x 18.1 mm.

## 2. Videos (associated with fig. 1c of main text):

For all supplementary videos, first the object was placed at a fixed distance (object distance) and the gap between the sensor and lens (image distance) was tuned to get the best possible image. After that, the object was moved back and forth and sideways for the video. For supplementary video 1 and 2, the object distance & image distance was roughly 300mm & 20mm, and 150mm & 22mm for video 3 and 4.

- Supplementary Video 1: Soldering iron (color palette: hot).
- Supplementary Video 2: Soldering iron (color palette: monochrome).
- Supplementary Video 3: Resistor (color palette: hot).
- Supplementary Video 4: Resistor (color palette: monochrome).

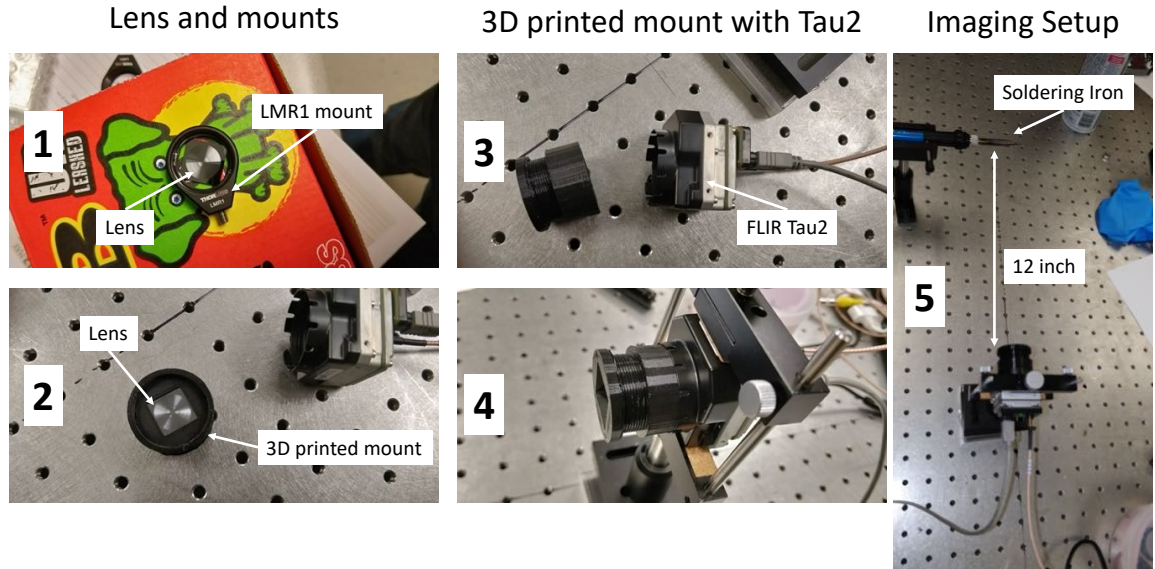

**Fig. S3:** Imaging test setup.

## 3. PSF Measurements

Collimation: The incident beam was collimated coming out of the QCL but expanded though a 3.3x beam expansion system to fill the MDL and ZnSe lens' apertures. We confirmed collimation using a NanoScan beam profiler to ensure that a focused beam FWHM was minimized over lens separation in z. Clear minimum had FWHM of 17um although diffraction limit of lens should have been 9um indicating focusing was not totally efficient and the PSF measurements were likely consistently larger than expected because of this.
PSF:
We placed the lens in the path of the expanded collimated beam and placed an FPA with 17um pixel size at its focus. We had to adjust the FPA distance in z until the spot size had been minimized. This was the PSF for the incident wavelength. We did this for all 4 wavelengths.

Towards camera

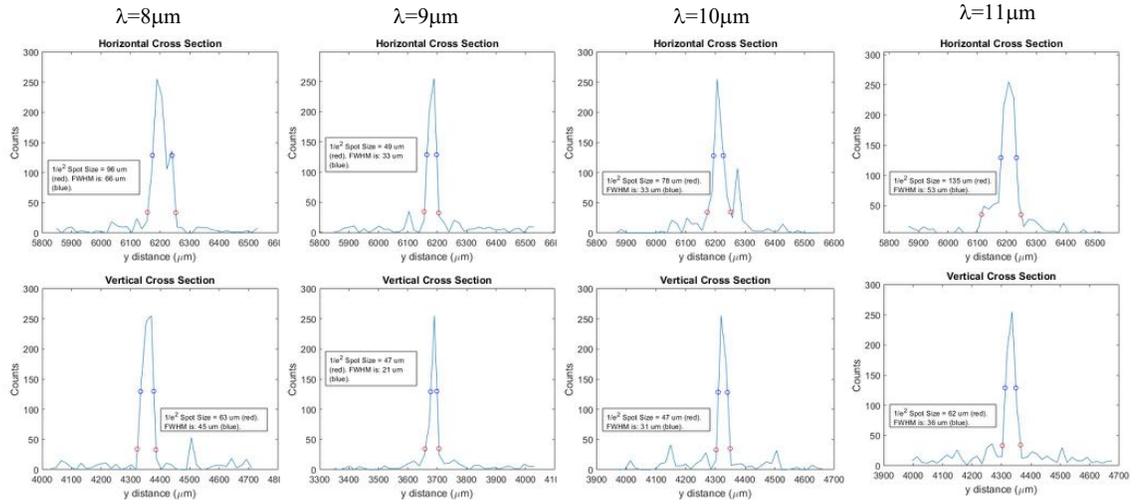

Towards object

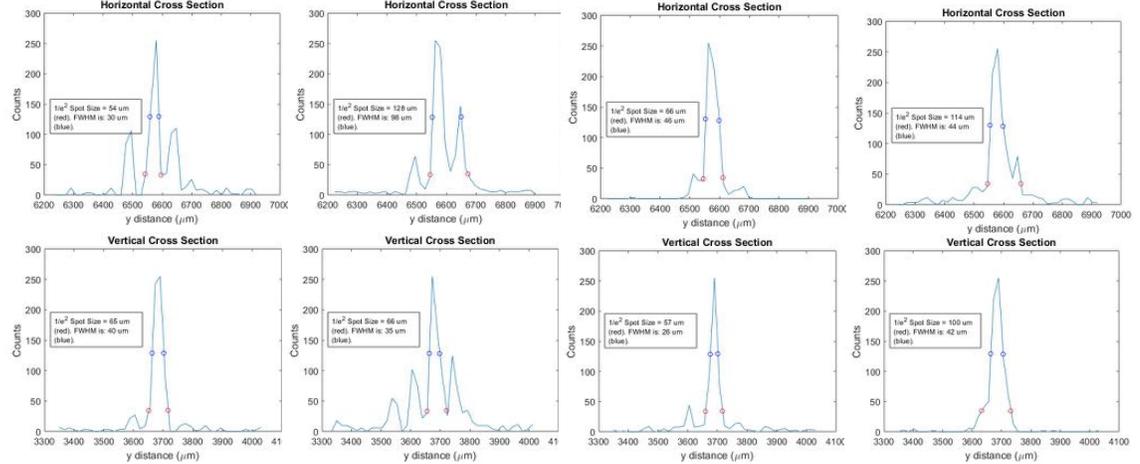

**Fig. S4:** Measured cross-sections of PSFs.

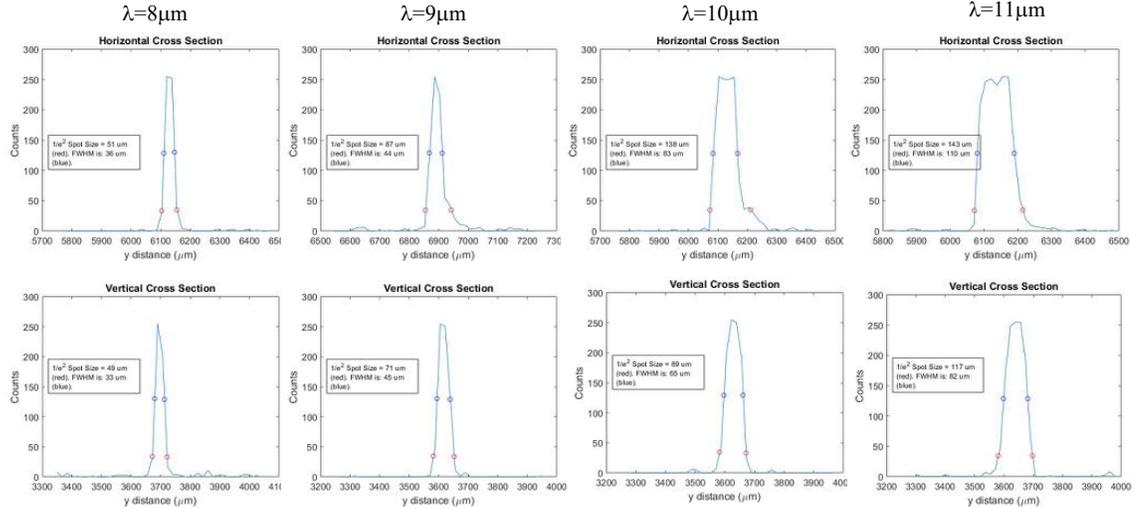

**Fig. S5:** Measured cross-sections of PSFs. Refractive Lens.

## 4. Varying the pulse width

We chose 9um for our wavelength of demonstration. We could control the incident power of the QCL by adjusting the pulse width and its rep rate. We chose 2 pulse width values that were an order of magnitude apart in length, but adjusted the rep rate so that the incident power of both was the same and took the PSF for both cases. We had to ensure that we did not saturate the FPA so the powers and hence the pulse widths were low.

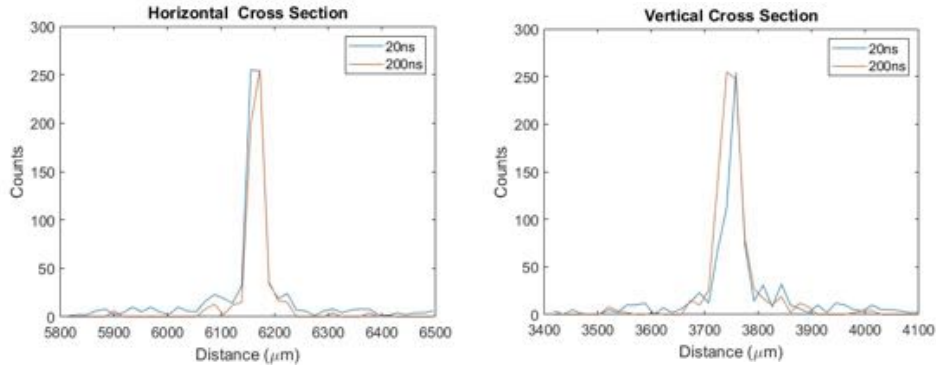

**Fig. S6:** Measured cross-sections of PSFs with varying pulse widths in the illumination beam.

## 5. Efficiency measurements

The beam's power is not distributed as a square, rather it is Gaussian. Furthermore, it is larger than the aperture of our power meter which means we will not be able to detect all the light incident on it. Therefore we will determine the power distribution and ration what we do measure to it. We took an iris and swept it over the beam in 1mm steps. We recoded the power at each step. This formed a P(x) curve which we then looked at to see how much power was contained within the middle 1cm. The ratio of the power within the middle 1cm and the total power was the ratio for the power incident.

We took the power measurement before the MDL (P1), immediately after (P2), and in the focal plane behind it (P3), and the same three for the refractive lens (see schematic in Fig. S8). In each cell of the following table is the MDL; Refractive Lens.

Table S1: Power efficiency measurements.

| Wavelength | P2/P1 = Transmission | | P3/P2 = Focusing | | P3/P1 = Total | |
|---|---|---|---|---|---|---|
| | MDL | Conv. | MDL | Conv. | MDL | Conv. |
| 8um | .0.22 | 0.86 | 0.71 | 1 | 0.15 | 0.86 |
| 9um | 0.19 | 0.91 | 0.73 | 0.98 | 0.14 | 0.89 |
| 10um | 0.25 | 0.9 | 0.77 | ~1 | 0.19 | 0.9 |
| 11um | 0.25 | 0.86 | 0.79 | ~1 | 0.19 | 0.86 |

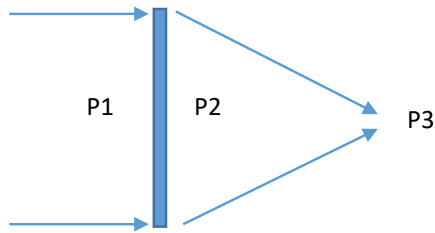

**Fig. S7:** Schematic indicating power measurements before (P1), right after (P2) the MDL, and at focal plane (P3).

## 6. Depth of Focus measurements

Choosing one wavelength, we adjusted the position of the FPA it in the z direction and took the PSF measurements. This allowed us to see how the focus changed in distance from the lens and to compare it to that of the conventional lens.

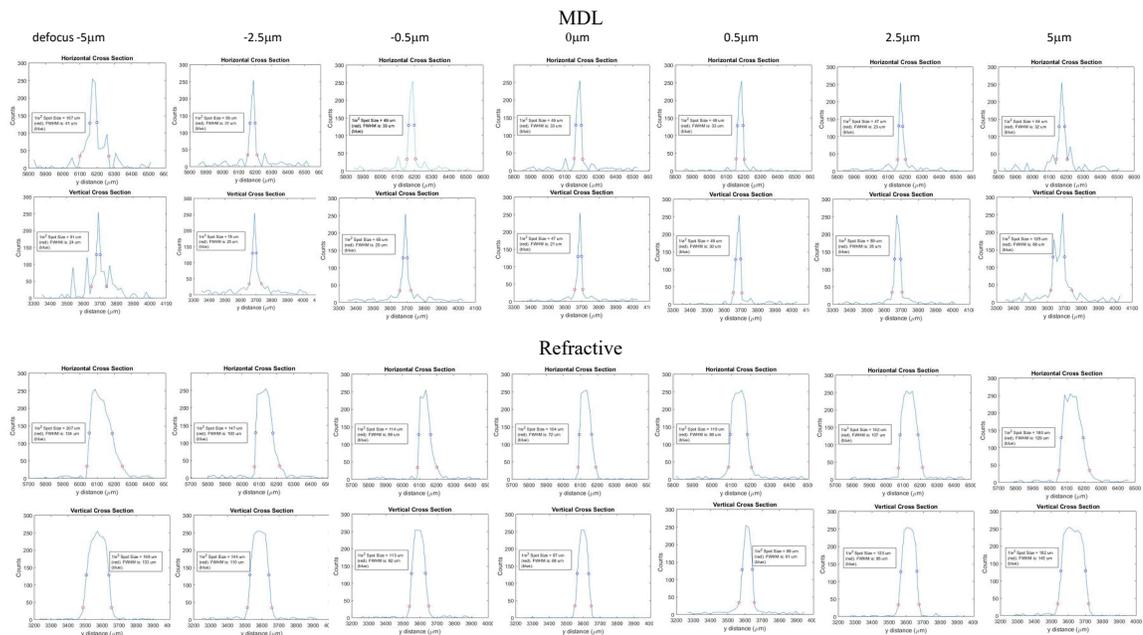

**Fig. S8:** Cross-sections of measured PSFs vs defocus for MDL and refractive lens.

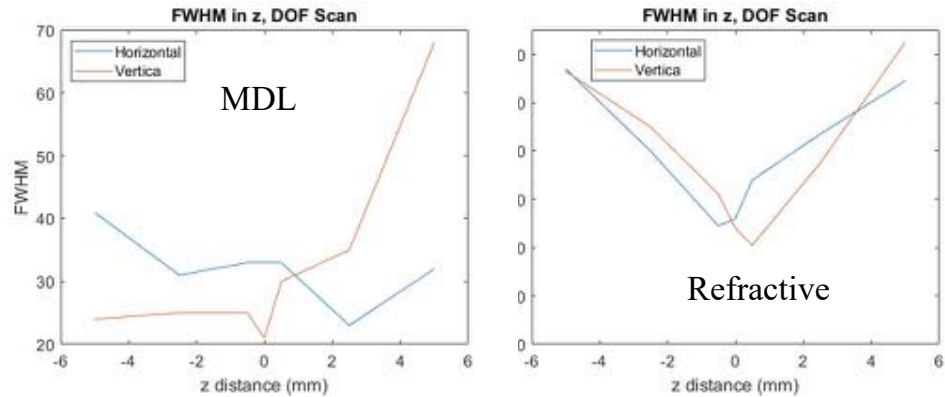

**Fig. S9:** Measured FWHM vs defocus for MDL and refractive lens.

## 7. Encircled Power:

We used a script to measure what percentage of the total power was concentrated in increasing radii within the MDL's PSF measured images. Summations of increasingly large concentering circles of the PSF plots were conducted and for each wavelength in different configurations and for the refractive lens. The sums were divided by the respective total intensities of the images in order to find the encircled power progression.

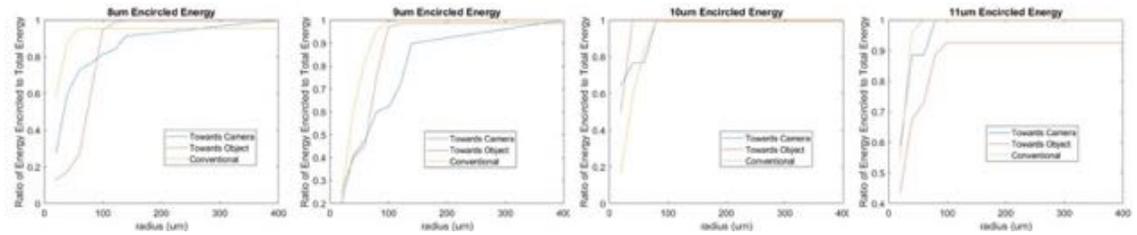

**Fig. S10:** Estimated encircled power.

## 8. Field of view measurements

Using a soldering iron as an illuminating object, the FPA and the object were situated 4 focal lengths apart with the MDL positioned in the middle, 2 focal lengths apart from both as illustrated in Fig. S11. This configuration ensured the object would be recreated on the FPA at the same size. Then, moving from left to right, the object measured intensity was recorded with the FPA moving so it was inverted. When the intensity at either side was half that of the intensity measured through the middle (the MDL's ideal object to image configuration), the FOV had ended.

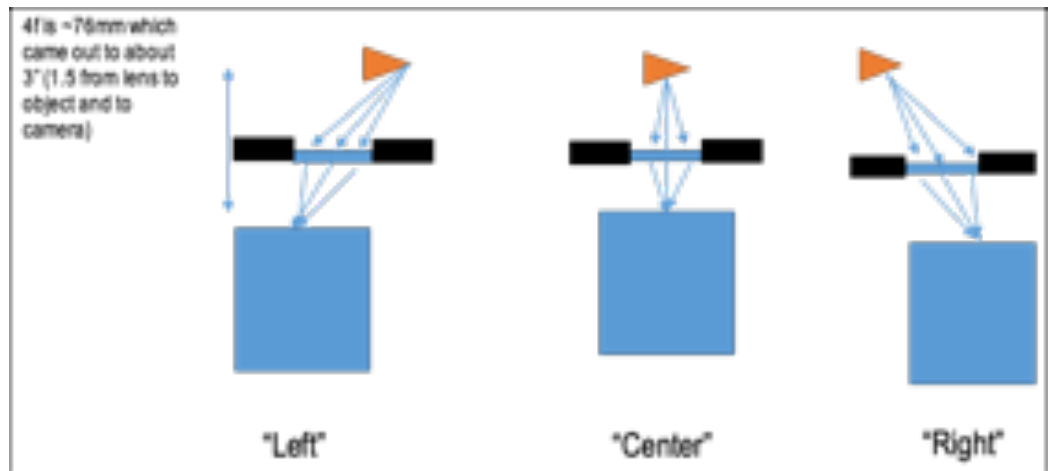

**Fig. S11:** Schematic showing FOV measurement.

We set up a heated soldering iron and an FPA, 4 focal lengths away from each other with the MDL in symmetric middle. We started them on opposite corners and moved them towards the middle and out again. We found where the intensity was half that of the peak for the declared edges of the FOV.

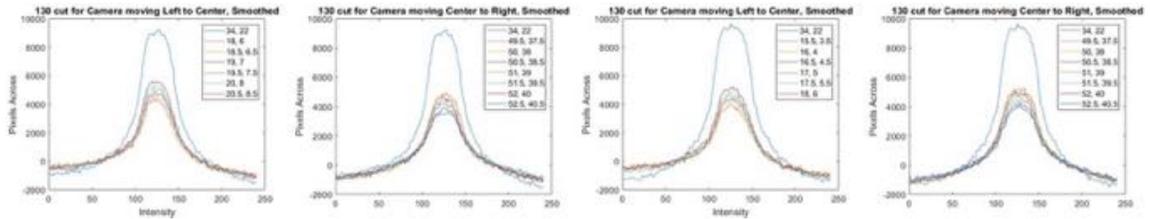

**Fig. S12:** Measurements to estimate FOV. Legend lists micrometer readings for given 130 intensity contour. Left 2 panels: towards camera. Right 2 panels: towards object.

## 9. Strehl Ratio measurements

Strehl Ratio is the ratio of the normalized measurement to its intensities in its theoretical Airy disk over the normalized ideal Bessel function pattern to its intensities within the theoretical Airy disk.

First the theoretical Airy disk is calculated $D = 2.44 * \lambda * f/\#$. All intensities in the measurement within this diameter are summed, then divide the measurement. A Bessel function is created around the maximum of the measurement. It too is normalized in the same procedure. The max of both normalized functions are ratioed (measured/ideal) for the Strehl ratio.

Table S2: Strehl ratio.

|      | MDL Facing Towards Camera | MDL Facing Towards Object | Conventional Facing Towards Camera |
|------|---------------------------|---------------------------|-----------------------------------|
| 8um  | .42                       | .34                       | .36                               |
| 9um  | .53                       | .48                       | .35                               |
| 10um | .5                        | .43                       | .32                               |
| 11um | .4                        | .41                       | .37                               |

## 10. Imaging measurements

We set up the FPA a focal length behind the MDL and placed various backlit images before it to see how well it resolved real objects. We also tried to push the limits on its ability to resolve gaps horizontally and keep objects in focus despite their horizontal separation. Images are visible in the body of the text.

## 11. MTF extraction from measured PSFs

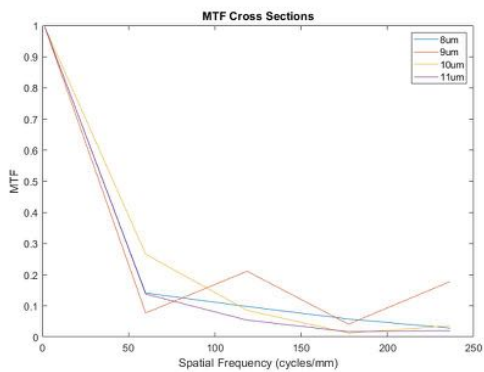

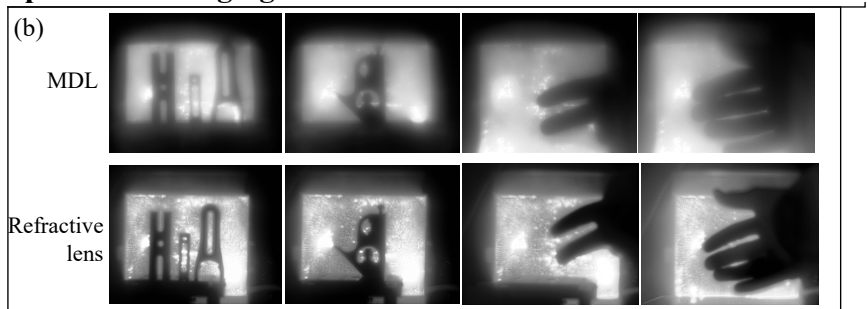

as well as the image with the source off.

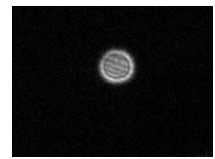

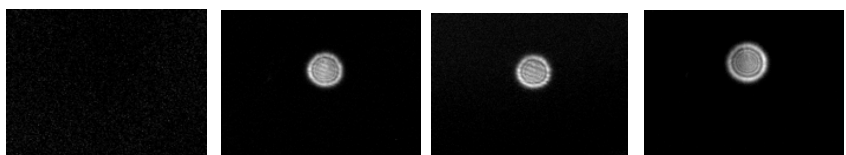

**Fig. S15:** Imaging through an iris alone.

12.